\documentclass[9pt,twocolumn,twoside]{pnas-new}

\templatetype{pnasresearcharticle}
\title{Circadian rhythm of dune-field activity}

\author[a]{Andrew Gunn}
\author[b]{Matt Wanker} 
\author[c]{Nicholas Lancaster}
\author[b]{Douglas A. Edmonds}
\author[d]{Ryan C. Ewing}
\author[a,e,1]{Douglas J. Jerolmack}

\affil[a]{Department of Earth and Environmental Sciences, University of Pennsylvania, Philadelphia, PA 19104, USA}
\affil[b]{Department of Earth and Atmospheric Sciences, Indiana University, Bloomington, IN 47405, USA}
\affil[c]{Division of Earth and Ecosystem Sciences, Desert Research Institute, Reno, NV 89512, USA}
\affil[d]{Department of Geology and Geophysics, Texas A\&M University, College Station, TX 77843, USA}
\affil[e]{Department of Mechanical Engineering and Applied Mechanics, University of Pennsylvania, Philadelphia, PA 19104, USA}

\leadauthor{Gunn} 

\significancestatement{Sand dune fields experience extreme daily temperature cycles. We document a consequent cycle of atmospheric instability that promotes strong afternoon winds that transport sand and dust. As a result, dune fields make (or modulate) their own surface winds. The influence of daily temperature cycles on surface winds increases with dune-field size, possibly aiding desertification. This effect should be enhanced on Mars, helping to explain sand transport despite the planet's thin atmosphere.}

\authorcontributions{Conceptualization, D.J.J., A.G., D.A.E. and R.C.E.; Methodology, all authors; Software, A.G.; Validation, all authors; Formal Analysis, A.G.; Investigation, A.G., M.W. and N.L.; Resources, all authors; Data Curation, A.G. and N.L.; Writing -- Original Draft, A.G.; Writing -- Review \& Editing, D.J.J., D.A.E., R.C.E., N.L. and A.G.; Visualization, A.G.; Supervision, D.J.J., D.A.E. and R.C.E.; Project Administration, all authors.}
\authordeclaration{The authors declare no competing interests.}
\correspondingauthor{\textsuperscript{2}To whom correspondence should be addressed. E-mail: sediment@sas.upenn.edu}


\begin{abstract}
Wind-blown sand dunes are both a consequence and a driver of climate dynamics; they arise under persistently dry and windy conditions, and are sometimes a source for airborne dust. Dune fields experience extreme daily changes in temperature, yet the role of atmospheric stability in driving sand transport and dust emission has not been established. Here we report on an unprecedented multi-scale field experiment at the White Sands Dune Field (New Mexico, USA), where we demonstrate that a daily rhythm of sand and dust transport arises from non-equilibrium atmospheric boundary layer convection. A global analysis of 45 dune fields confirms the connection between surface wind speed and diurnal temperature cycles, revealing an unrecognized climate feedback that may contribute to the growth of deserts on Earth and dune activity on Mars.
\end{abstract}

\dates{This manuscript was compiled on \today}
\doi{\url{www.pnas.org/cgi/doi/10.1073/pnas.XXXXXXXXXX}}

\begin{document}

\maketitle
\thispagestyle{firststyle}
\ifthenelse{\boolean{shortarticle}}{\ifthenelse{\boolean{singlecolumn}}{\abscontentformatted}{\abscontent}}{}

\dropcap{W}ind-blown sediments define the regional climate and topography of large swaths of Earth and other planets \cite{hayes2018dunes,lancaster2006linear,andreotti2009giant,jerolmack2012internal}. Sand saltates across deserts creating forms like dunes and ripples \cite{kok2012physics}, while dust lofted high into the atmosphere serves as a catalyst for other Earth-system processes \cite{lambert2008dust} like cloud nucleation \cite{demott2003african} and phytoplankton growth \cite{mahowald2005atmospheric}. Appropriate descriptions of sediment dynamics are important for revealing both past climate, through aeolian stratigraphic interpretations \cite{rodriguez2014archean}, and future climate, through dust in Earth-system models (ESMs) \cite{kok2012physics}. The latter is an especially important challenge considering predictions of increased aridity in Earth’s future \cite{berg2016land}. Current ESMs used to estimate dust and sand fluxes employ sediment transport algorithms to find the critical friction velocity required to initiate particle motion \cite{zender2003mineral}. A shortcoming of these algorithms is their neglect of atmospheric stability \cite{kok2012physics}, a property of the atmospheric boundary layer (ABL) important for momentum transfer \cite{garratt1994atmospheric,monin1954basic,dyer1974review}. Some ESMs use ABL schemes that account for stability; however, free-parameters in sediment transport algorithms are calibrated neglecting stability, resulting in biased predictions for global dust. Furthermore, all ABL schemes neglect the role of time-varying stability, instead opting for time-invariant alternatives \cite{dyer1974review,hu2010evaluation}, introducing another bias in predictions. Current methods can therefore be poor predictors of the conditions that lead to sediment transport, suggesting overall that we must better understand the bridge between atmospheric stability and wind-blown sand \& dust.

Friction velocity, $u_*$, is a parameter that represents the shear of wind impinging on a rough boundary \cite{garratt1994atmospheric}. In sediment transport studies, $u_*$ is usually derived by fitting time-averaged vertical profiles of horizontal wind velocity with the so-called Law of the Wall (or log law) theory \cite{sterk1998effect,martin2017wind}, which is derived under the assumptions of a steady and uniform, turbulent boundary-layer flow of an unstratified fluid. Although this theory may approximate flow in some portions of the Atmospheric Boundary Layer (ABL) sometimes, vertical density gradients create buoyancy forces that lead to large deviations \cite{garratt1994atmospheric,monin1954basic,dyer1974review,frank1994effects}. Empirical functions have been developed in studies of the ABL to characterize this deviation via the Monin-Obukhov similarity theory (MOST) \cite{garratt1994atmospheric,monin1954basic,dyer1974review}, which predicts enhanced $u_*$ aiding sand transport in unstably stratified conditions, and suppressed $u_*$ in stable conditions \cite{lanigan2016atmospheric}. MOST, however, assumes steady-state stratification and is therefore unable to capture the daily evolution of the boundary stress imparted by the ABL if there is sufficiently strong periodic solar heating \cite{cheng2005pathology}. Moreover, knowledge of $u_*$ is necessary but not sufficient to forecast sand and dust transport. Winds must exceed a critical shear velocity, $u_{*,cr}$, for transport and erosion to occur, and this threshold increases rapidly with soil moisture due to formation of liquid bridges \cite{nield2011aeolian,neuman2003effects,ravi2005field}. Fast and warm surface winds drive evaporation and a lower $u_{*,cr}$, while calm and cool conditions facilitate condensation and a rise in $u_{*,cr}$.

\begin{figure}
\centering
\includegraphics[width=0.95\linewidth]{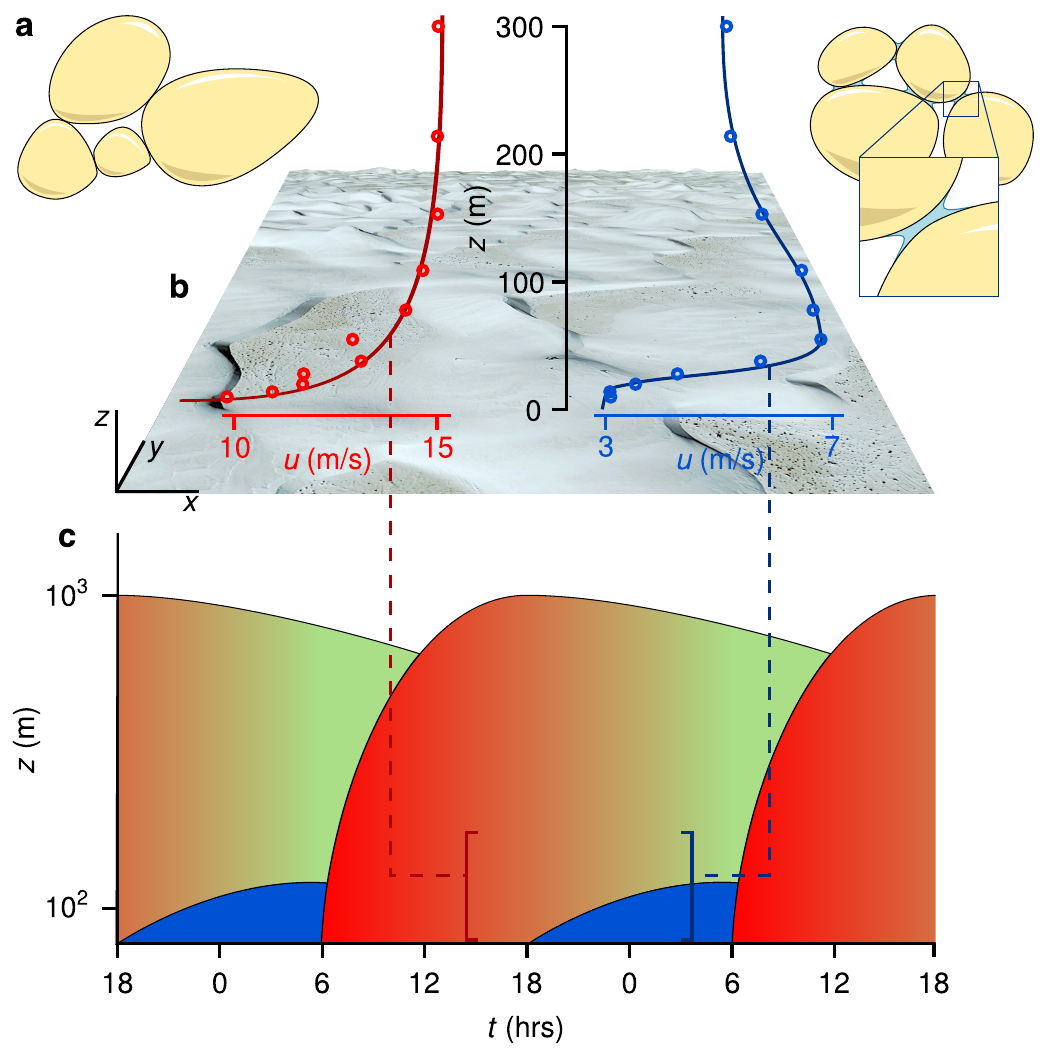}
\caption{Contrasting stability and humidity behavior of the ABL. (\textbf{a}) In the top corners, dry (left) and wet (right) grains are depicted aside wind profiles of corresponding ABL states. (\textbf{b}) Characteristic unstable (red) and stable (blue) wind speed ($u$) profiles are shown aloft White Sands (3x vertical extent) with raw instantaneous Doppler lidar data (circles; \textit{Materials and Methods}) from times below (dashed lines) overlaid. Note profiles are actually spatially co-located, but are shifted for visualization. (\textbf{c}) Schematic of daily ABL cycle; from sunset a stable layer (blue) grows slowly from the surface into the residual layer (green) which loses vertical extent, then at sunrise the surface warms initiating a rapidly growing unstable layer (red) until sunset, \textit{ad infinitum}.}
\label{fig:fig1}
\end{figure}

Considering the above dynamics, we present a unprecedented multi-scale dataset suggesting that a solar-driven daily cycle of wind, heat and humidity variation in the ABL can lead to a circadian rhythm of dune-field activity as follows (Fig. \ref{fig:fig1}). As the Sun warms the ground from sunrise into the day, that strong radiative flux evaporates water and heats air at the surface. This leads to thermal instability where near-surface vertical temperature gradients become out of equilibrium with the rest of the ABL. This departure from classical MOST, since the connection between momentum and heat throughout the column is lost, is due to the low thermal-inertia of the dune field. Convective forcing enhances turbulent mixing, eventually bringing momentum sourced from aloft down to the surface. As a result, surface winds speed up in the afternoon once the ABL has been fully mixed, by which point any surface moisture has been wicked away. Sand and dust transport are most likely to occur at this point, when surface winds are fast and the ground is dry. From sunset onward, surface air cools with the land faster than air aloft, slowly setting up stable thermal stratification \cite{garratt1994atmospheric,monin1954basic,dyer1974review} that blocks vertical mixing of momentum---and hence sediment transport from this mechanism---at night. Finally, night-time cooling eventually increases humidity and leads to condensation of water within surface sediments.

\section*{\textit{In situ} results}

\begin{figure}
\centering
\includegraphics[width=0.95\linewidth]{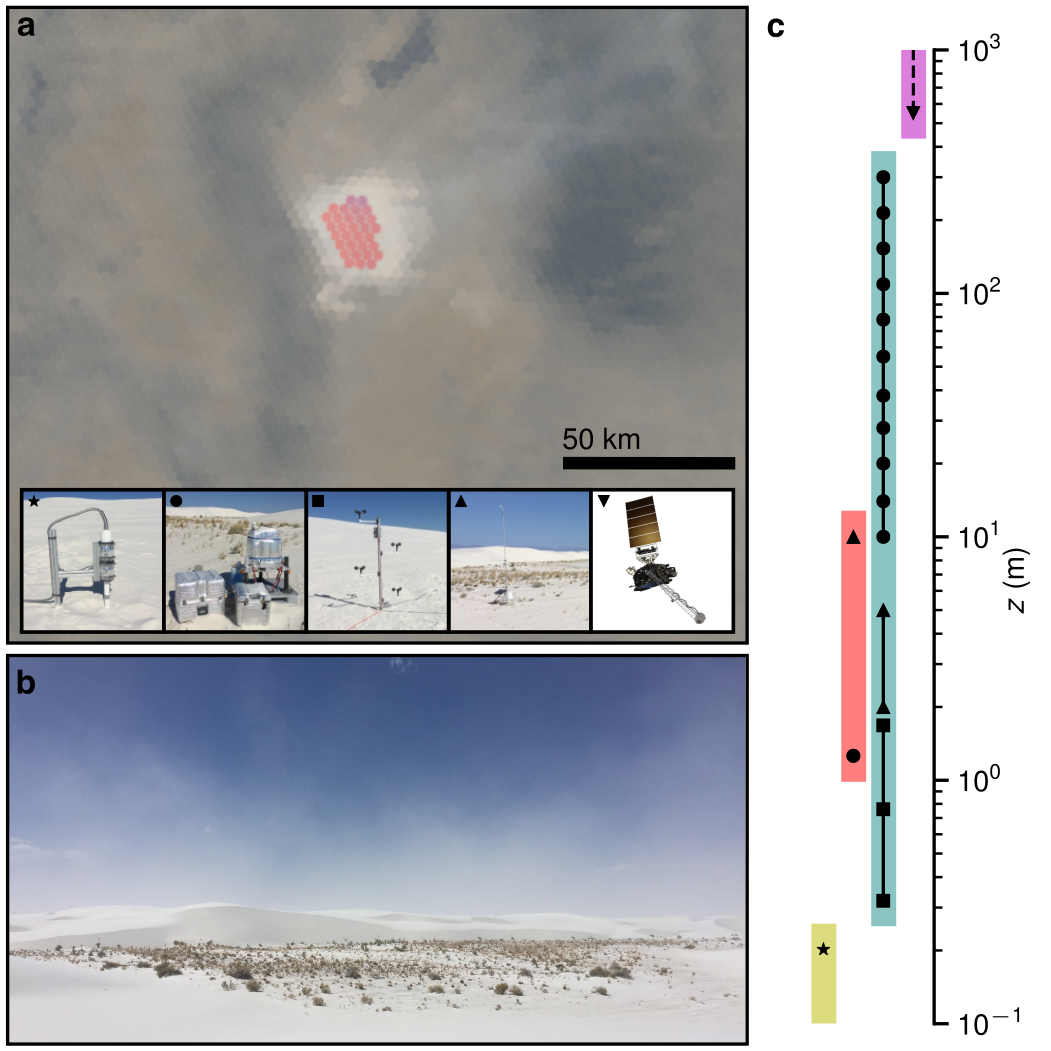}
\caption{FATE campaign systems and extent. (\textbf{a}) White Sands Dune Field captured in true color by the GOES-16 satellite during dust emission to the northeast, with RGB channels of the machine learning algorithm region predicting dust (red) overlaid. Inset, images of each system: Sensit H14-LIN sand saltation sensor (star); ZephIR 300 Doppler lidar wind-velocity profiler (circle); cup anemometer mast for near-surface wind speed (square); Met One meteorological tower for temperature, humidity and additional wind speed (up-triangle); and GOES-16 satellite for dust (down-triangle). (\textbf{b}) Transport activity at White Sands during campaign installation. (\textbf{c}) Elevation of sand saltation (yellow), atmospheric state (red), wind speed (turquoise) and dust emission (magenta); measurements are marked by symbols unique to each system.}
\label{fig:fig2}
\end{figure}

We report concurrent observations of the ABL, sand saltation and dust emission at the White Sands Dune Field (New Mexico, U.S.A.) that strongly support our hypothesis. For 24 days during the known March-April windy season associated with dune activity \cite{ewing2010aeolian}, we conducted a field campaign we call Field Aeolian Transport Events (FATE) (Fig. \ref{fig:fig2} and \textit{SI Appendix}, Fig. S1) that measured: sand saltation impacts; humidity; atmospheric stability; and horizontal wind speeds at 16 elevations \textit{in situ}. We developed a novel machine-learning dust-detection algorithm operating on images from the GOES-16 satellite (see \textit{Materials and Methods}, \textit{SI Appendix} Fig. S2 and Movie S1) to obtain synchronous synoptic-scale data on atmospheric dust. Together, these data provide an comprehensive view of the genesis of transport events that sculpt dunes and pump dust into the atmosphere at White Sands.

\begin{figure*}
\centering
\includegraphics[width=0.95\linewidth]{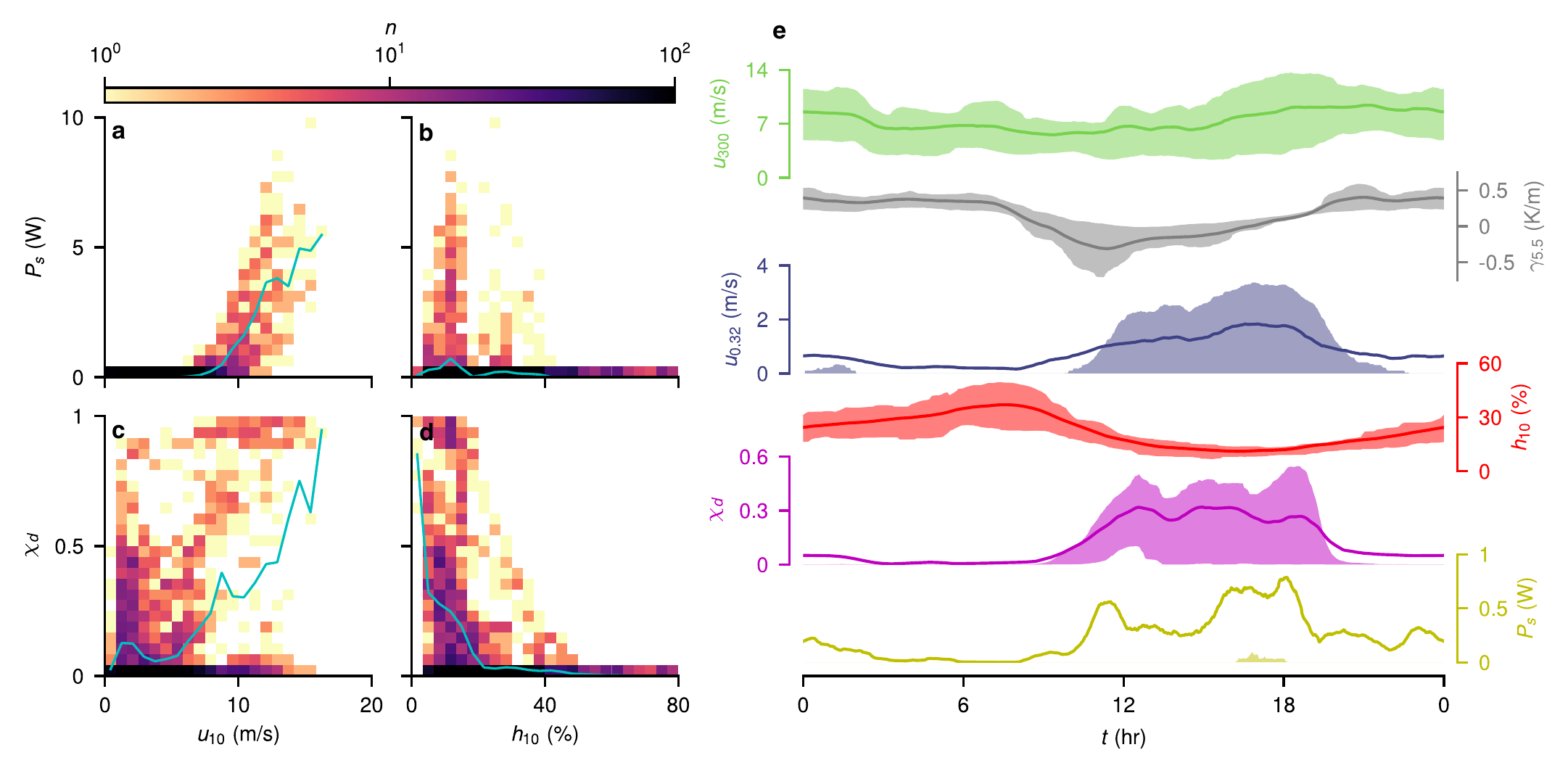}
\caption{Requisites for aeolian activity. 2D log-histograms of saltation power $P_S$ against concurrent (\textbf{a}) 10-m horizontal wind speed $u_{10}$ and (\textbf{b}) humidity $h_{10}$; (\textbf{c}) and (\textbf{d}) are similar histograms for dust suspension probability $\chi_{d}$. Mean horizontal value for each vertical bin is overlaid (cyan line). (\textbf{e}) The average 1-hr smoothed daily timeseries of horizontal wind speed at 300 m ($u_{300}$, green), potential virtual temperature lapse rate ($\gamma_{5.5}$, grey), horizontal wind speed at 0.32 m ($u_{0.32}$, blue), humidity at 10 m ($h_{10}$, red), dust probability ($\chi_{d}$, purple) and saltation power ($P_S$, yellow) (interquartile ranges are shaded).}
\label{fig:fig3}
\end{figure*}

Our data confirm previous findings \cite{kok2012physics,sterk1998effect,martin2017wind} that sand saltation is initiated around a threshold wind speed, and on average becomes more intense as winds pick up (Fig. \ref{fig:fig3}a). This behavior is variable, however, in part due to additional role of humidity. We find that the probability of occurrence of saltation diminishes with humidity; little to no transport occurs when relative humidity exceeds 40\% (Fig. \ref{fig:fig3}b). Similar trends are also observed for atmospheric dust (Fig \ref{fig:fig3}c-d), confirming the expected link between sand saltation and dust emission \cite{kok2012physics}. Having demonstrated high wind speed and low humidity at the surface indeed necessitate transport at White Sands, we now consider daily cycles in ABL dynamics that drive those surface conditions. In particular, we average data from all 24 days to produce a daily ‘climatology’ of the windy season. Our highest wind measurement is from 300 m elevation ($u_{300}$). The ABL thickness varies from $\sim10^2$ m to $\sim10^3$ m over a 24-hr period \cite{norton1976diurnal}; although this implies that $u_{300}$ is sometimes within the ABL, it is our closest approximation to the free-stream synoptic-controlled winds. The closest measurement to the surface occurs at 0.32 m ($u_{0.32}$). We see that on the average day, $u_{300}$ and $u_{0.32}$ do not co-vary (Fig. \ref{fig:fig3}e). Instead, high-elevation winds are fastest at night while low-elevation winds peak in the afternoon, a dynamic also observed in other arid landscapes \cite{hu2010evaluation}. We believe this is due to thermal instability. We quantify stability using the potential virtual temperature lapse rate at 5.5 m (see \textit{Materials and Methods}), $\gamma_{5.5}$, and find that peak negative stability occurs around the solar insolation maximum at noon, triggering enhanced surface winds that strengthen as the ABL is progressively mixed. Also in the afternoon, near-surface humidity reaches its daily minimum, indicating the desert surface is at its driest. As hypothesized, these dynamics culminate in sand saltation and dust emission that is focused in the afternoon (Fig. \ref{fig:fig3}e). At sunset (roughly 18:00) the atmosphere becomes positively stable, the surface winds and transport activity die off, and the upper atmospheric winds speed up.

These observations are hard to reconcile with the Law of the Wall theory, which is often used in sand and dust transport studies \cite{kok2012physics,sterk1998effect,martin2017wind}. Indeed, friction velocities derived from this method are poor predictors of saltation, although prediction is enhanced when MOST equations are used to compute $u_*$ (see \textit{Materials and Methods} and \textit{SI Appendix}, Fig. S3). Nonetheless, the best predictor of saltation is the surface wind velocity, as observed in other studies \cite{martin2017wind}. Although MOST explicitly incorporates atmospheric stability conditions, the steady-state assumption is broken in the strongly transient daily dynamics of the desert ABL (\textit{SI Appendix}, section 1, Figs. S4 \& S5).

One way to demonstrate violation of equilibrium is to examine the daily evolution of wind speed vs. stability as a trajectory in state-space (Fig. \ref{fig:fig4} and a longer Doppler lidar wind-velocity profiler deployment in 2017 shown in \textit{SI Appendix}, Fig. S6). An ABL with constant free-stream winds would have no path dependence, and hence a unique relation between the variables. We compare wind speed measured at three elevations---the lowest ($u_{0.32}$) and highest ($u_{300}$), and a popular reference ($u_{10}$)---to low-level ABL stability $\gamma_{5.5}$ (Fig. \ref{fig:fig4}a). It is clear that the ABL exhibits path dependence and hence a memory of state; winds at all elevations are slower at dawn than dusk for equal stability. We observe similar dynamics at White Sands on a nearby, low-roughness playa during an earlier field deployment in 2017 (\textit{SI Appendix}, Fig. S6). We find that such hysteresis does not arise for weakly-forced conditions, such as the authoritative CASES99 experiment \cite{poulos2002cases} (\textit{SI Appendix}, Fig. S7). The loops (Fig. \ref{fig:fig4}a) have an internal area because of this hysteresis; however, the loop skews toward a positive relation between stability and speed aloft, and is opposite for near-surface winds. This is because daytime instability ‘props open the door’ for momentum to be mixed down toward the ground, and nighttime stability closes it allowing a nocturnal jet to skim over the underlying cold air \cite{garratt1994atmospheric,hu2010evaluation,cheng2005pathology,poulos2002cases}. On average the fastest near-surface winds are not seen during the strongest instability, but actually at neutral stability. This is a consequence of the hysteresis: it takes time for thermal plumes to entrain free-stream momentum. Scaling $u_{300}$ and $u_{0.32}$ by their expected values derived from the Law of the Wall (see \textit{Materials and Methods}) highlights their distinctly opposing state-space trajectories (Fig. \ref{fig:fig4}b). This provides insight that the neutral-stability assumption breaks down at the most crucial time for sediment transport (when the surface ABL momentum is greatest) and, paradoxically, is most correct at times when buoyancy influence is extreme.

\section*{Global results}

We now generalize the insight from our field site by scrutinizing the diurnal ABL cycle over 45 dune fields during the past decade. We pair a newly constructed comprehensive atlas of active dune fields with a global hourly 32-km gridded reanalysis of meteorological observations from 2008-2017 derived from the ERA5 dataset \cite{copernicus2017era5} (see \textit{Materials and Methods} and \textit{SI Appendix}, Fig. S8). The magnitude of the diurnal near-surface temperature cycle is represented by the daily 2-m temperature range ($\delta T_2$), while the daily maximal 10-m wind speed ($u_{10,max}$) is our proxy for formative near-surface winds for sand transport that day. Dune-field size is represented by area ($A$), mapped from satellite data (see \textit{Materials and Methods}). As expected, we see that dune fields correspond to regions of the planet having the largest diurnal temperature ranges (Fig. \ref{fig:fig5}c).

\begin{figure}
\centering
\includegraphics[width=0.95\linewidth]{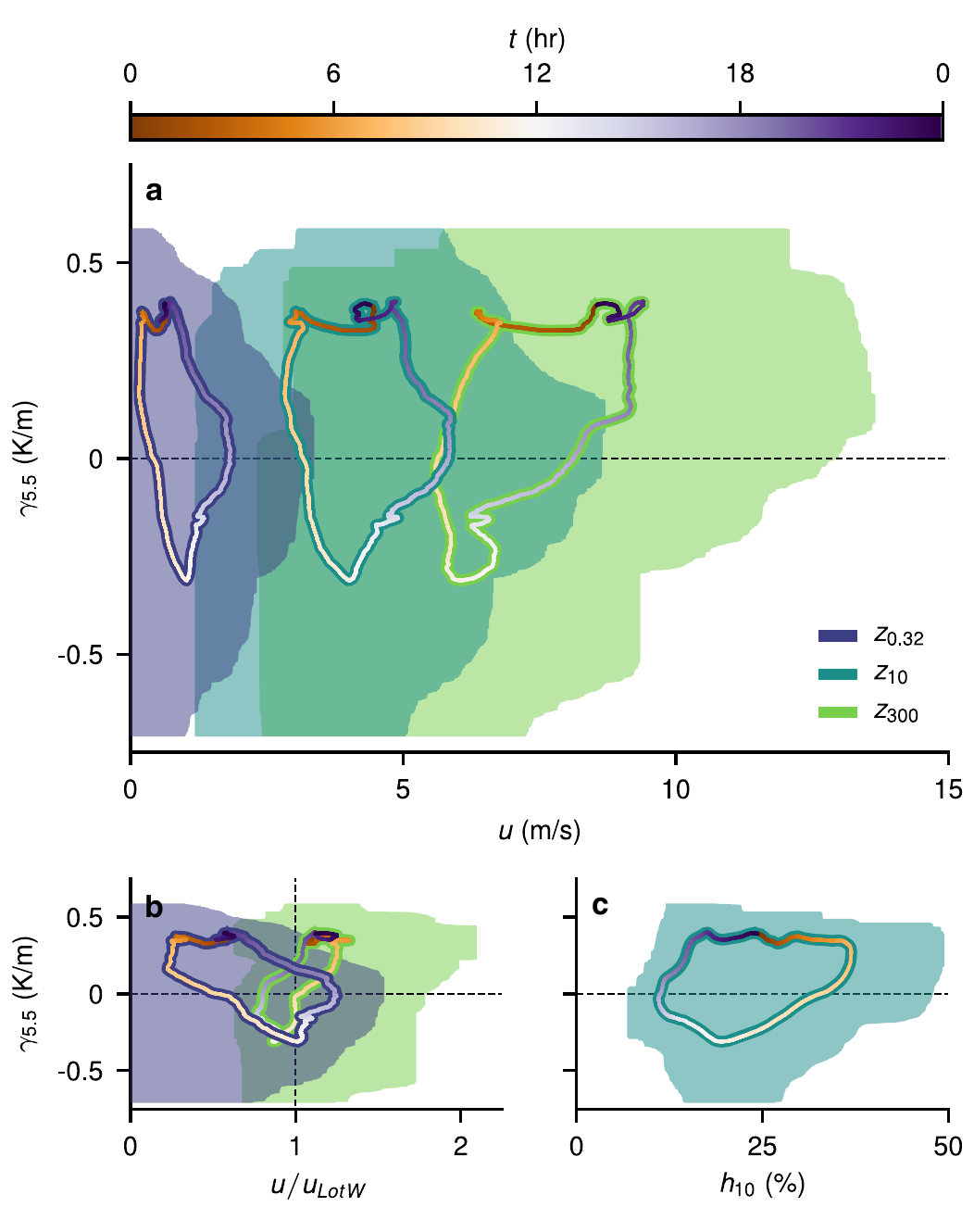}
\caption{Daily ABL trajectories. Each plot shows average 1-hr smoothed daily trajectories coloured by day-hour $t$ and outlined in solid color by measurement height (legend of (\textbf{a})). Shaded regions surrounding the trajectories are the union of interquartile range rectangles for all times. Trajectories are cast on a space spanned vertically by potential virtual lapse rate $\gamma_{5.5}$ and horizontally by; (a) horizontal speed $u$, (\textbf{b}) $u_{300}$ and $u_{0.32}$ scaled by a neutral Law of the Wall prediction $u_{LotW}$ given the observed $u_{10}$ and a roughness length of 0.1 m, and (c) relative humidity $h_{10}$.}
\label{fig:fig4}
\end{figure}

\begin{figure}
\centering
\includegraphics[width=1\linewidth]{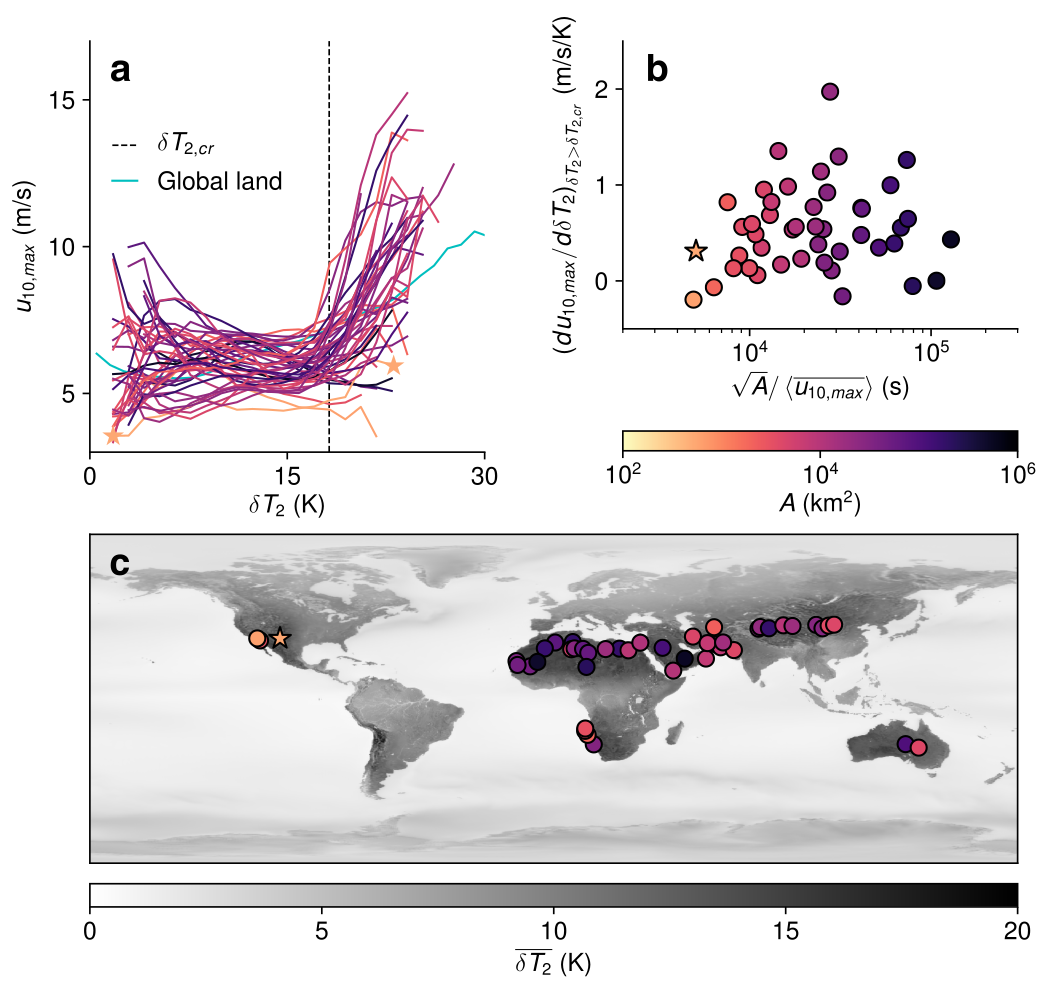}
\caption{Climate-land feedback via the aeolian diurnal cycle. (\textbf{a}) Average relationships between daily 2-m temperature range and 10-m maximum wind speed for each of the 45 dune fields examined (\textit{SI Appendix}, Fig. S8), coloured by area. The average for all global land (cyan) is shown for reference, and a critical daily temperature range $\delta T_{2,cr}$ (see \textit{Materials and Methods}) is marked (dashed line). (\textbf{b}) Characteristic atmospheric residence time is plotted against the slope of lines in (a) for $\delta T_2$ exceeding $\delta T_{2,cr}$ for each dune field. (\textbf{c}) Dune field locations marked on a map coloured by the time-averaged $\delta T_2$. In (b) and (c), markers are coloured by dune field area. White Sands is starred in all panels. Atmospheric data are from the ERA5 reanalysis of the decade 2008-2017 \cite{copernicus2017era5}}
\label{fig:fig5}
\end{figure}

The relation between $u_{10,max}$ and $\delta T_2$ for each dune field exhibits a striking threshold behavior (Fig. \ref{fig:fig5}a). Below a critical temperature range $\delta T_{2,cr} \approx 18$ K, the day’s fastest winds do not vary with changes in the daily heat cycle for any dune field. For strongly forced days where  $\delta T_2 > \delta T_{2,cr}$, however, fast winds are overwhelmingly positively correlated with the daily temperature range. We posit that this is a macroscopic signature of the onset of convective instability, a non-equilibrium phenomenon that arises only under sufficiently large thermal forcing. At White Sands, these dynamics correspond to observed wind profiles (Fig. \ref{fig:fig4}) that deviate from a steady-state description such as MOST \cite{cheng2005pathology}; we expect similar behavior at other dune fields, but \textit{in situ} measurements are lacking. We use a classical ABL model \cite{garratt1994atmospheric} to determine the maximum value for near-surface $\delta T$ at which a well-mixed characteristic ABL can maintain equilibrium. We find a critical value of $\delta T_{2,cr} = 18.1$ K, consistent with the proposed onset of non-equilibrium dynamics in the global data (see \textit{Materials and Methods} and \textit{SI Appendix}, Fig. S9).

A secondary trend in the global data is that the relation between $u_{10,max}$ and $\delta T_2$ appears to be stronger for larger dune fields beyond $\delta T_{2,cr}$ (Fig. \ref{fig:fig5}a). We hypothesize this is due to the ABL residence time over the dune field; the larger the dune field, the longer a column of air will experience its daily heat cycle. We compare the slope $K$ of the relation $u_{10,max} = K \delta T_2 + c$, for $\delta T_2 > \delta T_{2,cr}$, to a characteristic residence time of the ABL (Fig. \ref{fig:fig5}b). The overall positive trend indicates that longer residence times lead to maximal daily wind speeds that are more sensitive to diurnal temperature range. Therein lies a potential---and previously unidentified---positive climate-land feedback: larger sand seas create stronger winds by fostering non-equilibrium ABL dynamics through their high-amplitude daily temperature cycles (Fig. \ref{fig:fig5}c), in turn promoting dune activity and the outward migration of the sand-sea boundary. We attribute the scatter in this relationship, which grows with area, to coincident controls on dune field growth unique to each region \cite{kocurek1999aeolian,ewing2010aeolianbcs}; be they orographic, coastal, lithologic, biologic, tectonic or climatic ($\sim 10^4$ yr and greater). In short, we suggest that very large sand seas are more likely to be affected in their extent by sediment supply, basin boundaries, and long-term climate change.

\section*{Implications}

White Sands’ season of geomorphic work, when the most dust is emitted \cite{hinds1977boundary} and dunes migrate most \cite{ewing2010aeolian}, is also the driest and has the largest diurnal temperature range \cite{norton1976diurnal}. FATE provides a mechanistic explanation for this observation: both atmospheric instability and low humidity are necessary to initiate sediment transport. Our \textit{in situ} study is the first explicit demonstration of these dynamics, and the accompanying global study of many active dune fields on Earth serves to generalize them. Desert environments have the highest aridity and diurnal temperature ranges on Earth \cite{lancaster1984climate}. We posit that the dynamics reported here are not unique to the modern dune fields and time periods we studied, and point out that previous research has hinted at similar behavior in other dune fields \cite{frank1994effects,lanigan2016atmospheric,garcia2015turbulent}. Further, we suggest that dune fields nurture the ABL properties that lead to wind-blown sand, and that this coupling may strengthen as they grow, in-turn creating a positive feedback. In some cases this feedback may outpace stabilizing mechanisms for dune field boundaries, such as vegetation growth \cite{duran2006vegetation,reitz2010barchan}.

The understanding gained through this work may help to explain other phenomena. The positive feedback likely acts over geologic timescales in the expansion of sand seas, extending to the large dune systems during the Last Glacial Maximum and across subtropical supercontinents \cite{rodriguez2014archean}. We also expect the Martian ABL to adhere to these dynamics. Based on recent data \cite{millour2012mars,forget1999improved} we believe that Nili Patera, a well-studied and representative active dune field on Mars \cite{ayoub2014threshold}, has a daily near-surface temperature range approximately 25\% greater than $\delta T_{2,cr}$ during the dusty season \cite{banfield2020atmosphere} (see \textit{Materials and Methods}). We hypothesize that enhancement of surface winds by non-equilibrium ABL dynamics may help to resolve a current riddle on Mars, where surface winds simulated with MOST are too weak to explain actively migrating dunes \cite{newman2017winds,bridges2017martian}.

\matmethods{

\subsection*{Doppler lidar wind-velocity profiler}
The Zephir 300 machine measures the Doppler shift of a 1560 nm continuous-wave laser beam off passive tracers in the atmosphere. The beam is iteratively focused around distances with a lens as it traces a cone above using a revolving mirror, measuring averages of wind vectors centered within an Eulerian area at heights $z$ during revolution time \cite{wagner2009investigation}. After omitting revolutions with insufficient backscatter and power, the raw averages are linearly interpolated onto a grid at the most frequent raw timestep (17 s). The heights $z$ are chosen to distribute evenly in $\ln(z)$ between the machine’s maximum (300 m, set by a maximum lens probe volume at $d$), minimum (10 m, safety), including a mandatory (38 m) height. Temperature is measured on the machine at 1 m. The machine stores data locally and is powered by solar panels. Note that the lidar was also deployed for 70 days (March-May 2017) at White Sands in a different location at the playa-dune boundary, where surface roughness was much lower and concurrent measurements of other quantities were not made (\textit{SI Appendix}, Fig. S6). The locations (during FATE and in 2017), outside the influence of dune-steered flow in the formative direction, are shown in \textit{SI Appendix} Fig. S1.

\subsection*{Met tower}
This study uses measurements of humidity, pressure, temperature (all at 10 m) and wind speed (at 5 and 2 m) from a meteorological tower erected by the National Parks Service and Texas A\&M University marked in \textit{SI Appendix} Fig. S1. Data are output at 10-min timesteps as averages (set by local memory) using a Campbell Scientific CS1000 logger and linearly interpolated onto the same grid as the lidar for lapse rate calculations. The tower sends data by cellular modem (available as WHS02 at https://mesowest.utah.edu) and is powered by solar panels.

\subsection*{Cup anemometers}
We present data from 3 cups of a 4 cup anemometer mast (one cup had an electrical failure during the campaign) taking wind speed measurements at approximately log-spaced (0.32 m, 0.76 m and 1.68 m) heights. The cups are each calibrated with an optical gate to match the cup anemometers on the Met tower. Data are stored at 1 s timesteps (set by cup inertia). An Arduino stores data locally and is powered by solar panels. This small mast was erected adjacent to the lidar (location in \textit{SI Appendix} Fig. S1) to understand near-surface flow better. 

\subsection*{Saltation sensor}
The Sensit H14-LIN saltation sensor uses changes in resistivity of a crystal, housed in an aluminum cylinder, due to its deformation to infer the energy of sand grain impacts. The kinetic energy of all impacts within 10 s timesteps (set by local memory) are summed using a pulse height analyzer method, effectively ensuring no impact is uncounted. Mounted vertically with the instrument body above ground (Fig. 1), the crystal had an average height of 0.2 m above the surface of a barchan stoss during the deployment (\textit{SI Appendix}, Fig. S1). The sensor data are stored on a Campbell Scientific CS1000 logger and the system is powered by solar panels. Saltation events are defined as periods of time for which a smoothed power timeseries continuously exceeds the lower quartile of observed non-zero powers, with the smoothing timescale chosen as the integral timescale,
\begin{align*}
\tau=\int_0^T R_{PP}(t)dt
\end{align*}
Where $R_{PP}$ is the two-point correlation function of the unsmoothed power timeseries and $T$ is the campaign duration.

\subsection*{Satellite dust detection}
GOES-16 is a geostationary satellite with a radiometer imaging at 16 wavelengths at 5-min intervals. We downloaded L2 reflectance and brightness temperature data for the continental USA (available at https://registry.opendata.aws/noaa-goes/) during the deployment. Using true color reconstructions from the 3 shortest wavelengths, and a residual of the 8.5 $\mu$m and 11.2 $\mu$m wavelengths (revealing nocturnal clouds \cite{schmit2018applications}, we manually identify two time periods each of dust emission, clear sky and cloud cover over a region of White Sands (Movie S1), totaling 1.3\%, 3.2\% and 4.5\% (to 1 decimal place) of the deployment period, respectively. This constitutes a training set for a perceptron machine with 1 hidden layer of 100 rectified linear unit function nodes \cite{pedregosa2011scikit}, where wavelengths 3.9 $\mu$m, 8.5 $\mu$m, 10.35 $\mu$m and 11.2 $\mu$m are features, that classifies image pixels into dust (red), clear sky (blue) or cloud cover (green) with probability $\chi$. These features were used as they are non-zero throughout an entire 24-hr period, provide unique information and pertain to the low-atmosphere alone \cite{schmit2018applications}. Samples are passed to the machine scaled by a function that transforms each feature of the training set to have zero mean and unit variance. Each pixel at each timestep is then classified, where $\chi_d$ is the average probability of dust suspension for the region, and an event is defined as $\chi_d>0.5$. Without common methods of validation, we look to our concurrent saltation measurements to verify the algorithm, finding that during `certain' dust suspension ($\chi_d>0.95$) saltation occurs 77\% of the time, and during `certain' inactivity ($\chi_d<0.05$) saltation occurs 13\% of the time (\textit{SI Appendix}, Fig. S2). We believe mismatches are primarily due to (i) cloud coverage masking dust emissions, and (ii) that most dust from White Sands is from the playa upwind of the dunes \cite{white2015regional} and therefore not local to, or subject to the same sediment availability constraints as, the dunes where the saltation sensor is.

\subsection*{Potential virtual temperature lapse rate}
This quantity, $\gamma_{5.5}$, is the vertical gradient in virtual potential temperature between 1 m (at the Zephir 300) and 10 m (at the Met One tower). Virtual potential temperature is the temperature a parcel of air has when adiabatically transported to a reference pressure, defined as $\theta_v \equiv (1+a_1r_v)(T (p_{ref}/p)^{a_2(1+a_3r_v)})$ (where $r_v=a_4e/(p-e)$ is the mixing ratio, $e=a_5h10^{a_6(T+a_7)/T}$ is the vapor pressure, and $a_n$ are empirical constants defined at http://glossary.ametsoc.org/wiki/). Only the Met One tower measured pressure and humidity, so we assume they are identical at 1 m and 10 m in this calculation.

\subsection*{Friction velocity derivations}
The ABL flow parameters friction velocity $u_*$, roughness length $z_0$, Obukhov length $L$, and displacement height $d$ are derived separately for 10-min smoothed data for each measurement system. We smooth at this timescale because it is the shortest timescale of the longest timestep system, the met tower. For the cup anemometers, only neutral Law of the Wall fits without displacement height are performed because the measurements are not fully within the flow region where stability effects are noticeable, and we do not want to over-constrain the fit. We fit an $O(1)$ polynomial to the measurements in $\ln(z)$ to find $u_*$ and $z_0$ in $u=u_*\ln(z/z_0)/\kappa$. The same calculation is performed on the met tower measurements. For the Doppler lidar, a subregion of wind profiles are fitted to integral forms of the Monin-Obukhov similarity theory and a neutral Law of the Wall theory. The subregions are where speed monotonically increases with height (with a maximum height of 109 m) from the lowest (10 m) measurement, and there are 4 or more data points available. This definition ensures no over-constraint and application of the theory to the appropriate region of the ABL \cite{garratt1994atmospheric}. The integral form of MOST is given in Eqn. \ref{eqn:MOint}, 
\begin{figure*}[bt!]
\begin{align*}
    u = 
    \begin{cases}
          \frac{u_*}{\kappa}\left\{\vphantom{\frac12}\right.\ln\left(\frac{z-d}{z_0}\right)-2\ln\left(\frac{1+x}{2}\right)-\ln\left(\frac{1+x^2}{2}\right)+2\arctan(x)-\frac{\pi}{2} \left.\vphantom{\frac12}\right\}  & , \zeta<0\\
      \frac{u_*}{\kappa}\left\{\ln\left(\frac{z-d}{z_0}\right)+\beta\frac{z-d}{L}\right\} & , \zeta>0  \numberthis \label{eqn:MOint}
   \end{cases}
\end{align*}
\end{figure*}
where $x=(1-\gamma(z-d)/L)^{1/4}$, the stability parameter $\zeta=z/L$, $\beta=5$ and $\gamma=16$ \cite{dyer1974review}. Both definitions are fit to each profile and the one with lowest variance is chosen. The form of the neutral Law of the Wall theory (with displacement height) is $u=u_*\ln((z-d)/z_0)/\kappa$. All Doppler lidar fits are performed using a least-squares regression with a Cauchy loss function \cite{branch1999subspace}, starting on a landscape at typical values of flow parameters scaled by characteristic scale $\{u_*:4\cdot10^{-1},10^{-2}; z_0:10^{-1},10^{-2}; L:\pm10^3,10^1; d:10^0,10^{-1}\}$, respectively. Von Karman’s constant is $\kappa=0.38$ in all calculations. From this, we see that the nearest-surface horizontal wind speed is the best predictor of saltation flux.

\subsection*{Variance of flow parameters and saltation}
The coupling between ABL flow parameters, be it friction velocity or horizontal wind speed, and saltation power is calculated through the following routine that standardizes their different thresholds, magnitudes and sample sizes. Firstly, average saltation power values in bins of flow values (chosen such that there are 25 between the mean non-zero flow values and 0) are found. The bin where the average saltation power first exceeds 0.5 W is chosen as the threshold flow value for saltation (\textit{SI Appendix}, Fig. S3). Because threshold wind speeds and friction velocities from different measurements are all different, we then collapse the data by scaling to the threshold flow values and saltation power. Then, average scaled saltation powers $\hat{P_s}$ are found for all scaled flow values $\hat{u}$ when binned similarly (100 bins between 0 and 1). Finally, the average distance between the scaled data and its mean relationship with saltation power $F=\langle\hat{P_s}(\hat{u})\rangle$ is defined as the variance,
\begin{align*}
\sigma = \frac{1}{N}\sum_{i=1}^N|\left\{\hat{u},\hat{P}_s\right\}_i - \{\hat{u},F\}|
\end{align*}
Where $N$ is the total number of samples for the flow parameter. The lower the variance, the less scatter in the fit. The values are 0.0305 for $u_{0.32}$, 0.0309 for $u_{2}$, 0.0316 for $u_{10}$, 0.0454 for cup anemometer $u_*$, 0.0568 for met tower $u_*$, 0.0792 for netural Law of the Wall lidar $u_*$, and 0.0646 for Monin-Obukhov similarity theory derived lidar $u_*$.

\subsection*{CASES99 comparison}
The Cooperative Atmosphere-Surface Exchange Study (referred to in this paper as CASES99) was carried out to understand phenomena in the weakly-forced and primarily nocturnal atmospheric boundary layer. The location and timing of the experiment was chosen for clear and calm conditions over land \cite{poulos2002cases}. This leads us to expect conditions close to steady-state, acting as a useful counter-example to the strongly-forced, non-equilibrium dynamics observed at White Sands. Extensive measurements were taken to comprehensively document the boundary layer; in this study we only make use of a small subset of the data. Main tower high temporal resolution data from CSAT3 sonic anemometers was downloaded (https://data.eol.ucar.edu/) for October 1999 near Leon, Kansas \cite{poulos2002cases}. Prevailing wind speed and virtual temperature measurements were down-sampled from 20 Hz to 18 s temporal resolution using a box-car average to approximately match the doppler lidar at White Sands. 1.5 m and 10 m elevation virtual temperature measurements were converted to potential temperature $\theta_v \equiv T_v (p_{ref}/p)^{a_2(1+a_3r_v)}$ with $r_v$ fixed to the mean FATE value 2.33 g/kg, giving a gradient of potential virtual temperature lapse rate (\textit{SI Appendix}, Fig. S7c).

\subsection*{Global sand sea ERA5 reanalysis}
Polygons of 45 active sand seas with a distribution in area and latitude similar to all sand seas were created (\textit{SI Appendix}, Fig. S8). Each sand sea boundary was mapped as a shapefile using Google Earth to identify contiguous areas of dunes, recognizing that dune field boundaries are often not sharp. All polygons were drawn in the exact same manner on the same projection. Mapping was carried out at a coarse scale depending on the size of the sand sea, but always at a much finer resolution than the ERA5 grid spacing (see \textit{SI Appendix}, Fig. S8). Dune field naming is based on a combination of common-place convention in the scientific literature and the local naming, and these are chosen neutral to jurisdiction claims. A shoelace algorithm was used to find the area of each sand sea.

The ERA5 reanalysis dataset \cite{copernicus2017era5} was downloaded (https://registry.opendata.aws/ecmwf-era5/), giving hourly 2-m temperature and 10-m prevailing wind speed over the decade 2008-2017. These data are gridded at 32-km resolution from a GCM strongly constrained by many forms of observation \cite{copernicus2017era5}. Data for each sand sea is derived from land grid cells from the ERA5 that overlap with their respective polygons. Mean relationships between daily maximum 10-m winds and 2-m temperature change are found using averages of the former in 35 bins between 0 K and 40 K of the latter (Fig. \ref{fig:fig5}a). Bins including less than 0.01\% of the total data for a given sand sea are excluded from the relationship. In total the sand seas analyzed in this study account for 1.63\% of the global land data used, and each day of the decade for each land grid cell totals $9.9\cdot10^7$ points.

\subsection*{Critical diurnal temperature range}
We employ characteristic scales for parameters in the so-called `slab model' of well-mixed ABL growth \cite{garratt1994atmospheric} to estimate a critical diurnal temperature range for which the ABL evolution transfers from being in equilibrium to out-of-equilibrium with surface heating (Fig. \ref{fig:fig5}a). The change in the average potential temperature in the mixed layer is modeled as,
\begin{align*}
\frac{d\langle\theta\rangle}{dt} = \frac{1+A}{z_i}b_0
\end{align*}
where $A$ is the so-called Ball parameter \cite{ball1960control}, $z_i$ is the layer height, and $b_0$ is the heat flux at the surface. Assuming a dry mixed layer, often the case in sand seas \cite{norton1976diurnal,lancaster1984climate}, so $\theta=T+\Gamma_dz$, the near-surface air temperature ($T$) evolution is,
\begin{align*}
\frac{dT}{dt} = \frac{1+A}{z_i}b_0 - \frac{1}{2}\frac{d\langle\Gamma_dz_i\rangle}{dt} \numberthis \label{eqn:delt1}
\end{align*}
If we assume a sinusoidal near-surface air temperature evolution during the day, $T=\delta T\sin(2\pi t/\tau_D)/2+\bar{T}$, then the maximum daily change in $T$ is,
\begin{align*}
\max{\frac{dT}{dt}} = \frac{\pi\delta T}{\tau_D}\numberthis \label{eqn:delt2}
\end{align*}
Substituting Eqn. \ref{eqn:delt2} into \ref{eqn:delt1}, and employing characteristic scales (including $dz_i/dt \sim z_i/(\tau_D/2)$) for this equation, we find $\delta T = ((1+A)\tau_Db/H_{ABL}-\Gamma_dH_{ABL})/\pi$ when $T$ is changing most. Assuming $\delta T$ is approximately $\delta T_2$ a critical range of $\delta T_{2,cr}=18.1$ K occurs for typical values of $A=0.2$, $\tau_D=1$ day, $F_{Hs}=550$ W/m$^2$ (where $F_{Hs} = b_0\rho c_p$, $\rho=1.2$ kg/m$^3$ and $c_p=1004$ J/kg/K), $H_{ABL}=1$ km, and $\Gamma_d=-9.8$ K/km. See \textit{SI Appendix} Fig. S9 for a sensitivity diagram of $\delta T_{cr}$ to $F_{Hs}$ and $H_{ABL}$.

\subsection*{Nili Patera, Mars, ABL calculation}
The dune field Nili Patera is located at {8$^{\circ}$N, 67$^{\circ}$W} and has active dune migration, especially in the `dusty season' (Solar Longitude $L_s$ of 270±15$^{\circ}$) on Mars \cite{ayoub2014threshold}. Using the Mars Climate Database \cite{millour2012mars,forget1999improved}, the most common GCM dataset used in studies of Mars' atmosphere that is validated using observations, we found values for 4 of the free parameters in our characteristic equation for the critical daily near-surface temperature range, $\delta T = ((1+A)\tau_DF_{Hs}/H_{ABL}-\Gamma_dH_{ABL})/\pi$ (see the \textit{Materials and Methods} section above for a derivation). The values at Nili Patera ($L_s=270$) are also quite representative of the Martian ABL in general: $F_{Hs}=85$ W/m$^2$, $\rho=0.015$ kg/m$^3$, $c_p=770$ J/kg/K, $H_{ABL}=6$ km. With $A=0.2$ and $\tau_D=88775$ s fixed, this leaves the dry adiabatic lapse rate, which is estimated as $\Gamma_d=-4.3$ K/km elsewhere \cite{leovy2001weather,banfield2020atmosphere}. The resultant critical range is $\delta T_{2,cr}=49.8$ K. With the Mars Climate Database \cite{millour2012mars,forget1999improved} predicting a diurnal temperature range of $\delta T_2 = 62$ K at Nili Patera ($L_s=270$), similar to measurements at Bagnold Dune Field \cite{bridges2017martian}, we find $\delta T_2 = 1.25 \cdot \delta T_{2,cr}$, a value that sits comfortably in the observed range of above-critical $\delta T_2$ on Earth.

\subsection*{Data Availability}
In situ data and all code from this study are publicly available at https://github.com/algunn/FATE. GOES-16 satellite data are available at https://registry.opendata.aws/noaa-goes/. ERA-5 reanalysis data are available at https://registry.opendata.aws/ecmwf-era5/.

}

\showmatmethods{}

\acknow{We thank David Bustos and White Sands National Monument for field support, Scott R. David for field assistance, and Keaton Cheffer for constructing equipment. Funding provided by; National Science Foundation NRI INT award \#1734355 to D.J.J.; White Sands National Monument through NPS-GC-CESU Cooperative Agreement \#P12AC51051 to R.C.E; and International Society of Aeolian Research through the Elsevier Aeolian Research Scholarship to A.G..}

\showacknow{}

\bibliography{pnas-sample}

\end{document}


\maketitle

\SItext

\section{Monin-Obukhov Similarity Theory}

The similarity solutions for the atmospheric boundary layer (ABL) proposed by Monin \& Obukhov \cite{monin1954basic} (MOST) give a unique solution to bulk quantities and profiles of wind speed $u(z)$ (m/s) and potential temperature $\theta(z)$ (K) with values for the Obukhov length $L$ (m), friction velocity $u_*$ (m/s), roughness length $z_0$ (m) and the surface temperature $\theta_0$ (K) \cite{garratt1994atmospheric}. The latter can be replaced with some other thermal information, say the surface buoyancy flux $B_0$ (m$^2$/s$^3$) or sensible heat flux $F_{Hs}$ (W/m$^2$). The form of these solutions has to be found empirically \cite{dyer1974review}, and formally the (ABL) should be in a steady-state for them to apply. For weakly-transient ABLs, MOST works well because the timescale for local properties of heat and momentum to impact the entire surface-layer is shorter than the timescale over which boundary conditions are forcing them to change. This constraint can be violated in many circumstances \cite{cheng2005pathology}; it can be expected when buoyancy fluxes change rapidly for example. In subsection 1.A, we show that when momentum and heat fluxes are free to vary independently, hysteric behavior qualitatively similar to what is observed at White Sands can exist using MOST. In subsection 1.B, we assess the validity of MOST using the data from White Sands to the best of our ability using non-ideal parameters. For comparison, a like assessment is also carried out on the CASES99 experiment, a dataset known to fit MOST well \cite{poulos2002cases}.

\subsection{Theory}

The equations below are summarized graphically in Fig. S4. At White Sands during a regular day, the near-surface wind speed $u_{10}$ (m/s) lags the near-surface bulk potential temperature gradient $\gamma_{5.5}$ (K/m) (Fig. S7). This implies that locally the buoyancy $b$ is not co-varying with the momentum since,

\begin{align*}
    \Psi_M(z/L) &= 
     \begin{cases}
         2\ln\left(\frac{1+x}{2}\right)+\ln\left(\frac{1+x^2}{2}\right)-2\arctan(x)+\frac{\pi}{2} & , \zeta<0\\
         -\beta\frac{z}{L} & , 
         \zeta>0  
         \numberthis \label{eqn:MOint_m}
    \end{cases},\\
    \Psi_H(z/L) &= 
    \begin{cases}
        2\ln\left(\frac{1+x^2}{2}\right)  & , 
        \zeta<0\\
        -\beta\frac{z}{L} & , 
        \zeta>0
        \numberthis \label{eqn:MOint_h}
   \end{cases},\\
\end{align*}

\begin{align*}
  u(z) &= \frac{u_*}{\kappa}\left[\ln\left(\frac{z}{z_0}\right)-\Psi_M(z/L)\right], \numberthis \label{eqn:MOint_u} \\
  \theta(z) &= \theta_0 + \frac{b}{u_*\kappa}\left[\ln\left(\frac{z}{\alpha z_0}\right)-\Psi_H(z/L)\right], \numberthis \label{eqn:MOint_theta}
\end{align*}

where $x=(1-\gamma z/L)^{1/4}$, $\alpha = 7.4$, $\beta = 5$, $\gamma = 16$, $\kappa = 0.4$ \cite{dyer1974review}. Near-surface buoyancy and wind speed need not co-vary, but they will if the total momentum flux of the ABL is constant (i.e. there is steady free-stream wind). Here we produce this effect explicitly by imposing the following time-varying heat flux $b_0$ and wind speed $u_{10}$,

\begin{align*}
    b_0(t) &= 
    \begin{cases}
          A_b\sin(2\pi t)  & , 0<t<\frac{1}{2}\\
          \frac{9A_b}{4\pi}\left[\left(4(t-\frac{1}{2})-1\right)^8-1\right] & , \frac{1}{2}<t<1  \numberthis \label{eqn:bt}
   \end{cases},\\
    u_{10}(t) &= 
    \begin{cases}
          B_u+\frac{9A_u}{4\pi}\left[\left(4(t-\tau+\frac{1}{2})-1\right)^8-1\right] & , 
          0<t<\tau\\
          B_u+A_u\sin\left(2\pi (t-\tau)\right)  & ,
          \tau<t<\tau+\frac{1}{2}\\
          B_u+\frac{9A_u}{4\pi}\left[\left(4(t-\tau-\frac{1}{2})-1\right)^8-1\right]& , \tau+\frac{1}{2}<t<1
          \numberthis \label{eqn:ut}
   \end{cases},
\end{align*}

where $A_b$ is the peak positive heat flux, $A_u+B_u$ is the peak wind speed $\tau$ afterward, and $B_u$ sets the nighttime wind, because $t$ is measured in days with $t=0$ is sunrise. The form of these functions is arbitrary, especially for the wind, but ensures that; the functions are relatively smooth, net heat flux over the day is 0, and the shape of the heat flux time-series is realistic. Now with $\theta_0(t) = B_\theta+ A_\theta\sin\left(2\pi(t-1/8)\right)$, Eqns. \ref{eqn:MOint_u}-\ref{eqn:ut} can be solved numerically with feasible values for the constants. Examples are shown in Fig. S4, where panel b shows results similar to Fig. S7 and $F_{Hs}\equiv b_0c_p\rho$, where $\rho$ is density and $c_p$ is the specific heat.

This demonstration shows that despite the CASES99 experiment have virtually no hysteresis in Fig. S7c, this does not imply that MOST cannot account for hysteresis. What it does show, is that to create this effect with MOST (Fig. S4b) there must be something other than the buoyancy flux driving the near-surface winds. This is not what we see at White Sands (Fig. 3e), and furthermore within the framework of the most common empirical forms of MOST (Eqns \ref{eqn:MOint_u} \& \ref{eqn:MOint_theta}), the nocturnal behavior isn't reproducible.

\subsection*{Observations}

Identifying explicitly how MOST breaks down at White Sands is unfortunately not possible, as we did not have high temporal-resolution measurements to directly compute turbulent fluxes. Taking a different approach within the scope of our measurements, we compute the buoyancy flux $B_0$ using two alternative methods and compare the results. In and of itself this exercise is useful, but we also compare CASES99 and FATE, since CASES99 is known to behave well with MOST treatment albeit in a more stable regime.

By fitting 30-minute average wind profiles to Eqn. \ref{eqn:MOint_u} with the free parameters $L$ and $u_*$, we attain $B_0\equiv -u_*^3/\kappa L$. We call this prediction $B_{0,M}$ as it is computed solely from the momentum of the ABL. The profiles are taken for winds at 7 elevations between 1 and 55 m for FATE (with $z_0=10^{-1}$) and 1.5 and 50 m for CASES99 (with $z_0=10^{-4}$). Then using measurements of $\theta_{1}$ ($\theta_{1.5}$ for CASES99) and $\theta_{10}$, we use the following assumption to attain $B_{0,H}$,
\begin{align*}
    B_0 &= C_Hu\frac{g}{\theta_0}(\theta_0-\theta),\\
    C_H &= \frac{\kappa^2}{\left[\ln\left(\frac{z}{z_0}\right)-\Psi_M(z/L)\right]\left[\ln\left(\frac{z}{\alpha z_0}\right)-\Psi_H(z/L)\right]},
\end{align*}
which is done using the $L$ derived from the momentum profile fit and without $\theta_0$ since we have two $\theta$ measurements above. If the ABL is `well-behaved' with respect to MOST, then these two methods should should produce very similar results at expected magnitudes ($B_0<F_{Hs,max}g/\rho/c_p/\theta_{0,min}\approx0.02$ m$^2$/s$^3$).

The former method produces more reasonable values of $B_0$, despite both methods suffering from the numerically challenging form of Eqn. \ref{eqn:MOint_m} (Fig. S5a,b). We see that for FATE the 30-min averages, some profiles produce extremely small Obukhov length fits from the momentum profile in the unstable limit (Fig. S5b inset), and some wind profiles tend to predict spuriously large buoyancy fluxes (Fig. S5b). Compared to CASES99, the mismatch between the $B_0$ predictions is larger for reasonable values (Fig. S5c). Overall this data suggests that MOST cannot capture the full extent of the unstable White Sands ABL, however explicit measurements are required to say this definitively.

\begin{figure}
\centering
\includegraphics[width=0.486\textwidth]{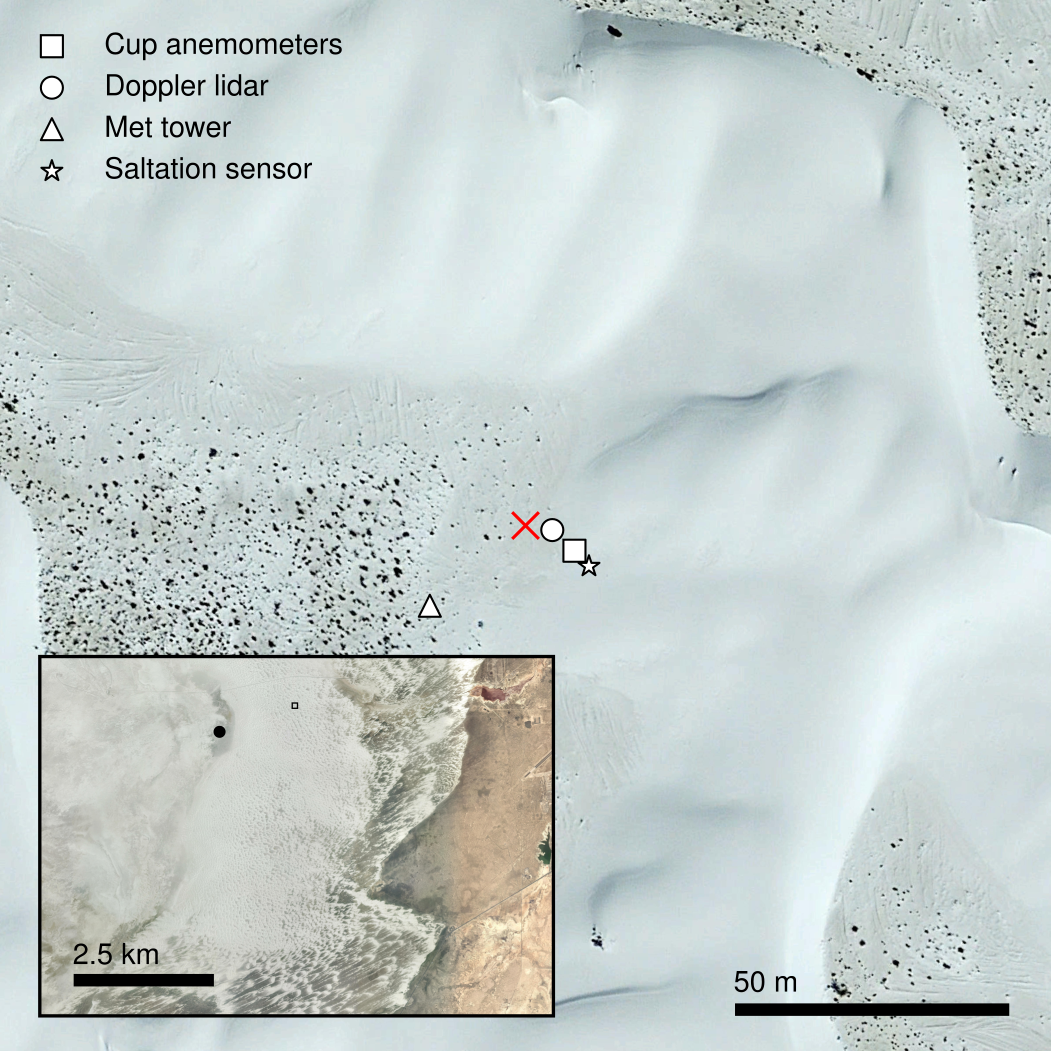}
\caption{Locations of \textit{in situ} measurement systems at White Sands. In the main image, symbols are centered on the locations of the cup anemometers (square), Doppler lidar wind-velocity profiler (circle), met tower (up triangle) and saltation sensor (star). Inset is a wider image of White Sands dune field, with a square at top-center indicating the main image location, and the location of the 2017 lidar deployment (black circle). Scale bars in both images show horizontal extent, and the red cross marks $\{32^{\circ}52'32.15''N, 106^{\circ}15'7.65''W\}$.}
\end{figure}

\begin{figure}
\centering
\includegraphics[width=0.487\textwidth]{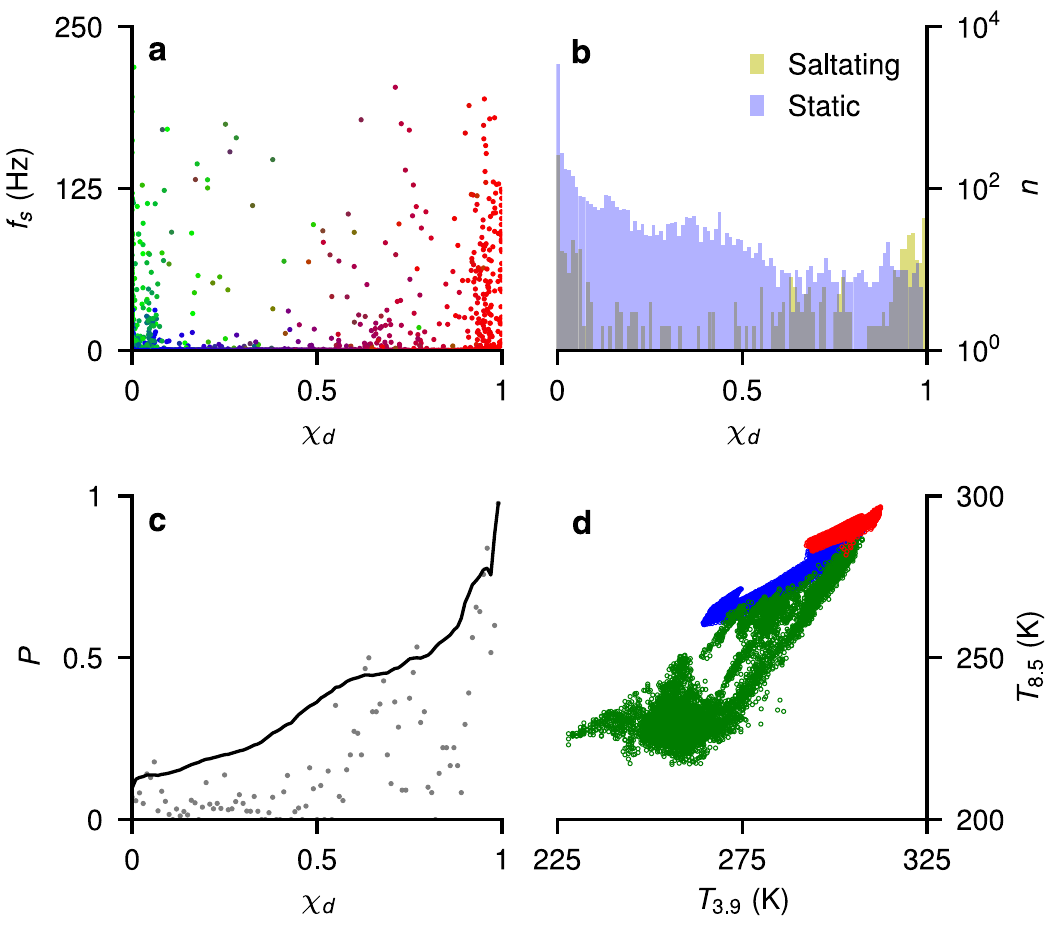}
\caption{Dust detection algorithm validation at White Sands. (\textbf{a}) Average probability of dust for the region plotted against saltation sensor collision frequency, where circles are colored in RGB (dust, cloud and clear sky, respectively). (\textbf{b}) Log-histograms of average probability of dust for the region for periods of a saltating or static sand surface measured by the saltation sensor. (\textbf{c}) Probability that the dust detection algorithm predicted saltation in 0.01 intervals (grey circles) or cumulatively (black line). (\textbf{d}) The training set of dust (red), cloud (green) and clear sky (blue) pixels in the region for two of the features (wavelengths 3.9 $\mu$m and 8.5 $\mu$m).}
\end{figure}

\begin{figure}
\centering
\includegraphics[width=1\textwidth]{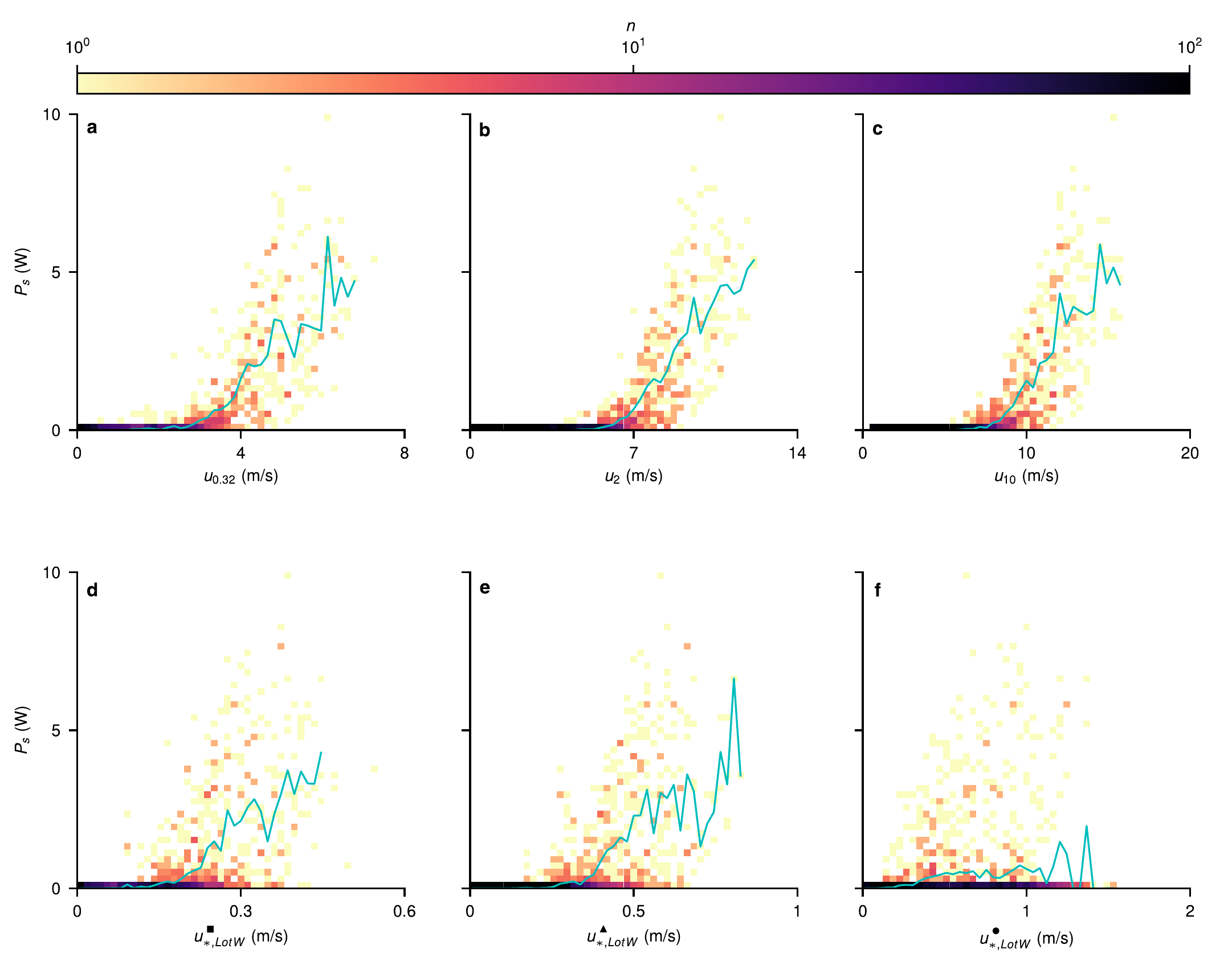}
\caption{Fluid and sediment transport coupling at White Sands. 2D log-histograms of 10-min smoothed saltation power and a measure of ABL flow are overlaid by the mean saltation power for each horizontal histogram bin (cyan line). The top row is the bottom speed measurement from each \textit{in situ} device, and the bottom row is a classic neutral Law of the Wall derived friction velocity. The left column is from the cup anemometers, middle from the met tower, and right is from the lidar. The relationship between the parameters degrades from (\textbf{a}) to (\textbf{e}) (see \textit{Materials and Methods}), and threshold friction-velocity changes dependent on observation height (\textbf{d}-\textbf{f}).}
\end{figure}

\begin{figure}
\centering
\includegraphics[width=0.7\textwidth]{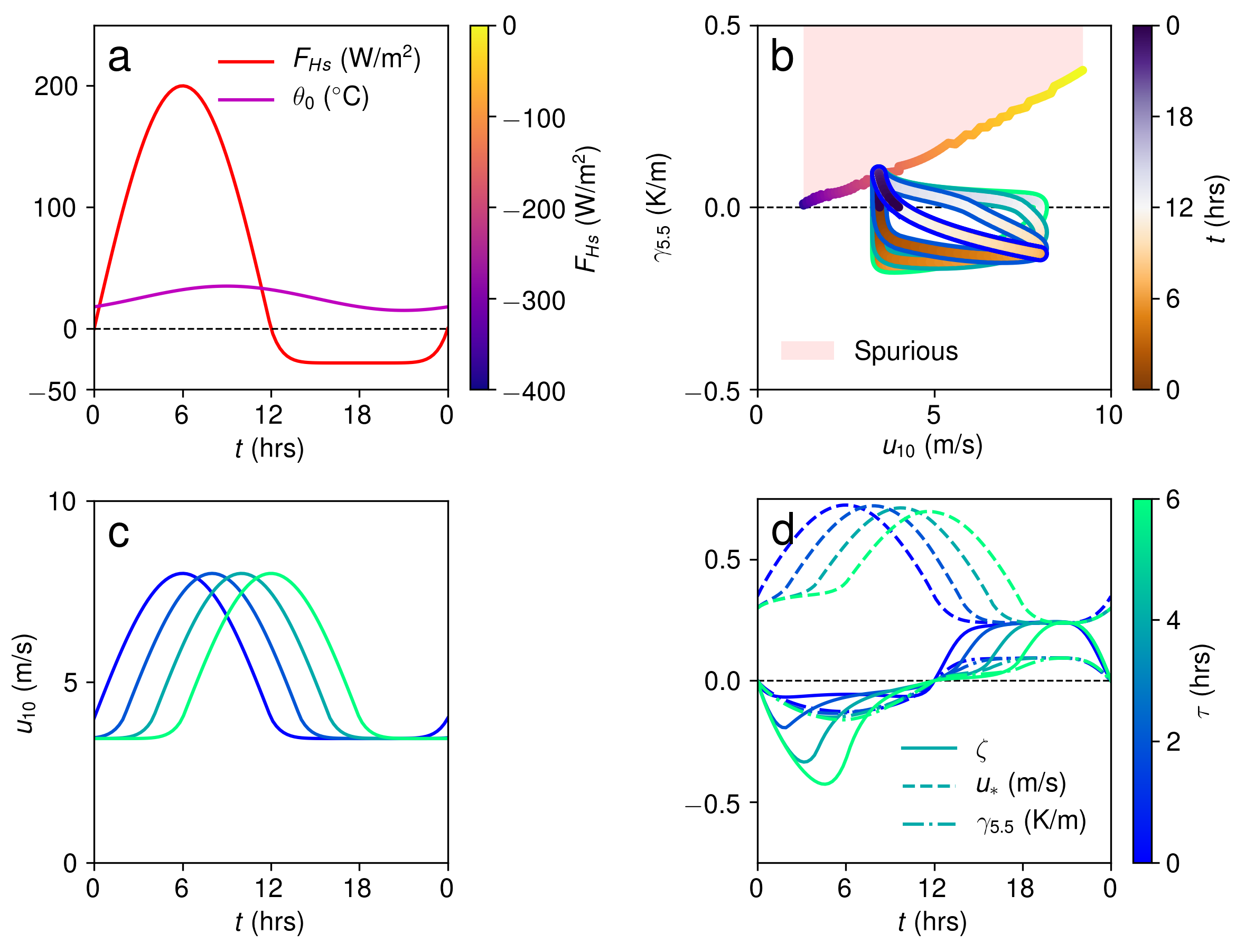}
\caption{Theoretical daily ABL cycles using Monin-Obukhov Similarity Theory. Hypothetical sensible heat flux $F_{Hs}$ (W/m$^2$) (a), surface potential temperature $\theta_0$ (K), and 10-m wind speed $u_{10}$ (m/s) daily cycles (equations are given for these in section 1). In (c) four scenarios for the time between the peak in heat flux and peak in wind speed $\tau$ (hrs) are given. (b) The resultant trajectories of wind speed and bulk gradient in virtual potential temperature $\gamma_{5.5}$ (K/m) are given for each lag time $\tau$ scenario. Without any lag there is no hysteresis in (b), and hysteresis grows with $\tau$. In (b) using standard solutions \cite{dyer1974review} for MOST, the area shaded red is numerically unstable, and the value of $F_{Hs}$ (W/m$^2$) at the boundary is given in the colorbar of (a). (d) The resultant values of the stability parameter $\zeta=10/L$ (where $L$ (m) is the Obukhov length), friction velocity $u_*$ (m/s) and $\gamma_{5.5}$ are given.}
\end{figure}

\begin{figure}
\centering
\includegraphics[width=1\textwidth]{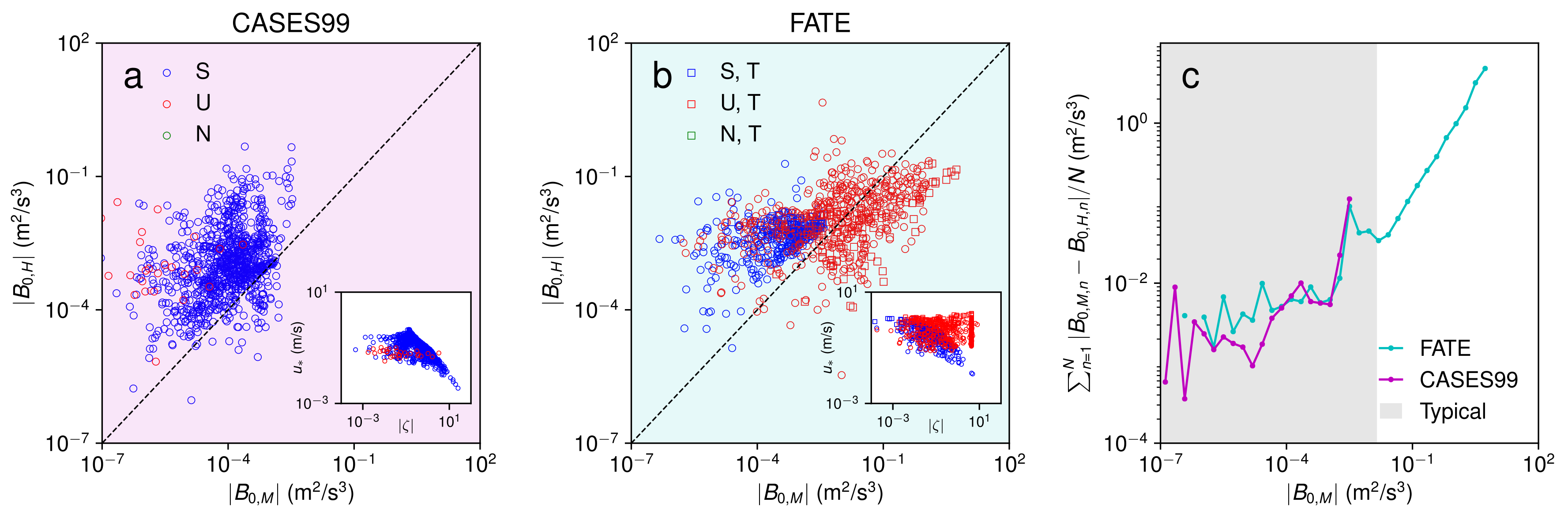}
\caption{Applicability of Monin-Obukhov Similarity Theory to CASES99 and FATE. In (a) and (b), buoyancy flux computed in two separate ways (see section 1 for an explanation) is shown for the CASES99 \cite{poulos2002cases} experiment (a) and the data reported here (FATE) (b). The background color of these plots matches the legend in (c), the insets are the corresponding values for $\zeta=10/L$ (where $L$ (m) is the Obukhov length) and the friction velocity $u_*$ (m/s) from the wind profile fit, and the legend shows the color of the points correspond to stable (S), unstable (U) and neutral (N) best fits to wind profiles. Profiles during transport (T) are shown as squares. (c) A measure of how different the buoyancy flux predictions are for a given buoyancy flux. The vertical axis shows the mean absolute difference between predictions within a bin on the horizontal axis. Within the typical range of values (grey shaded region) the FATE results are less similar between the methods than CASES99, and unreasonably large values exist for the FATE results also, implying the profiles are not consistent with MOST.}
\end{figure}

\begin{figure}
\centering
\includegraphics[width=0.487\textwidth]{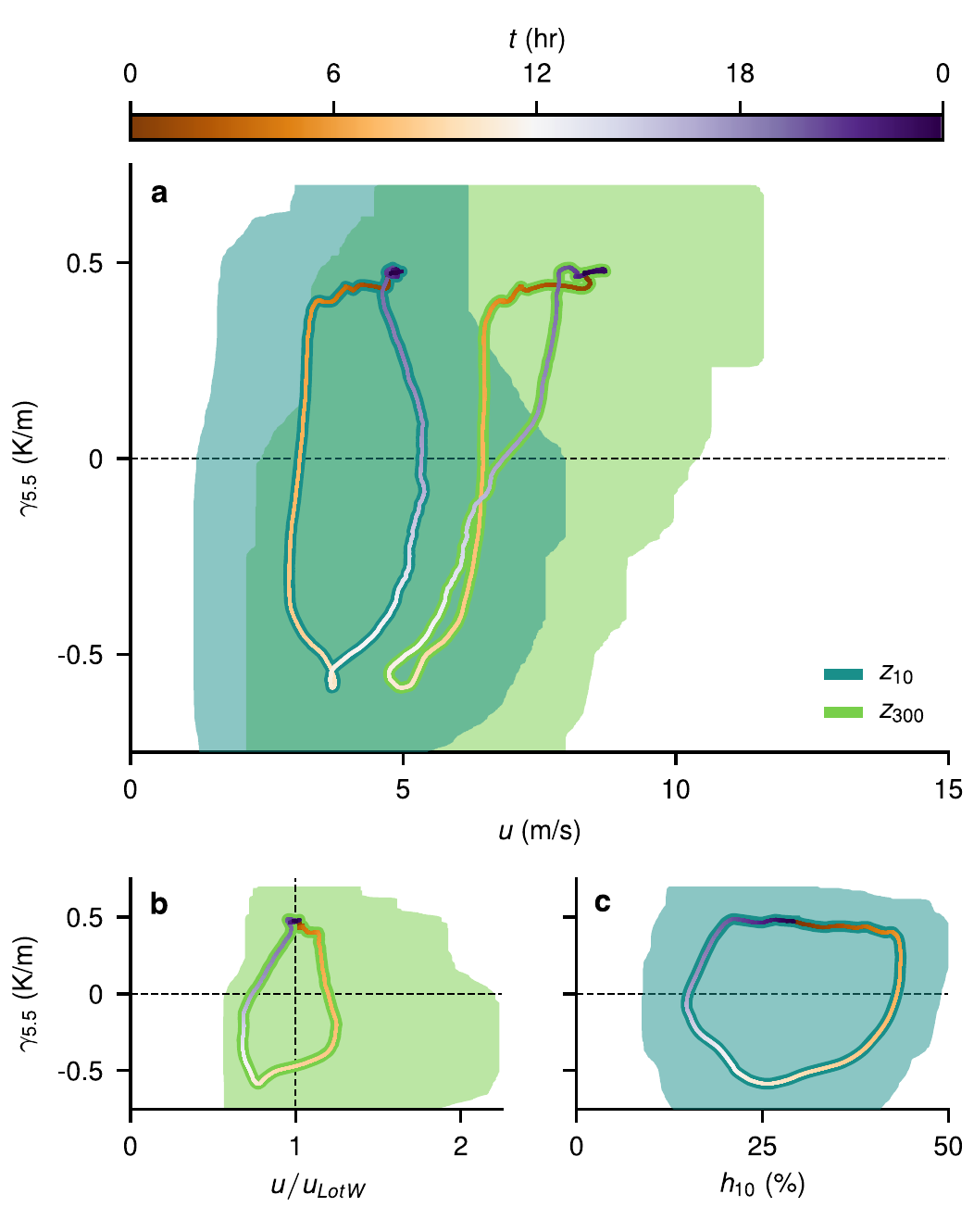}
\caption{Daily ABL trajectories at White Sands derived from a 70-day campaign in 2017. Data taken from a different location several kilometers away, and just upwind, of the beginning of the dune field on a flat deflation surface known as the Alkali Flat (Fig. S1). This figure is produced in a similar manner to Fig. 4, however only uses Doppler lidar wind-velocity profiler data (no $u_{0.32}$ measurements were taken). The dynamics here are qualitatively similar to that in Fig. 4. Differences may be due to variations in atmospheric conditions from year to year, the change in sampling duration used to generate daily climatology, and also the difference in boundary roughness.}
\end{figure}

\begin{figure}
\centering
\includegraphics[width=0.952\textwidth]{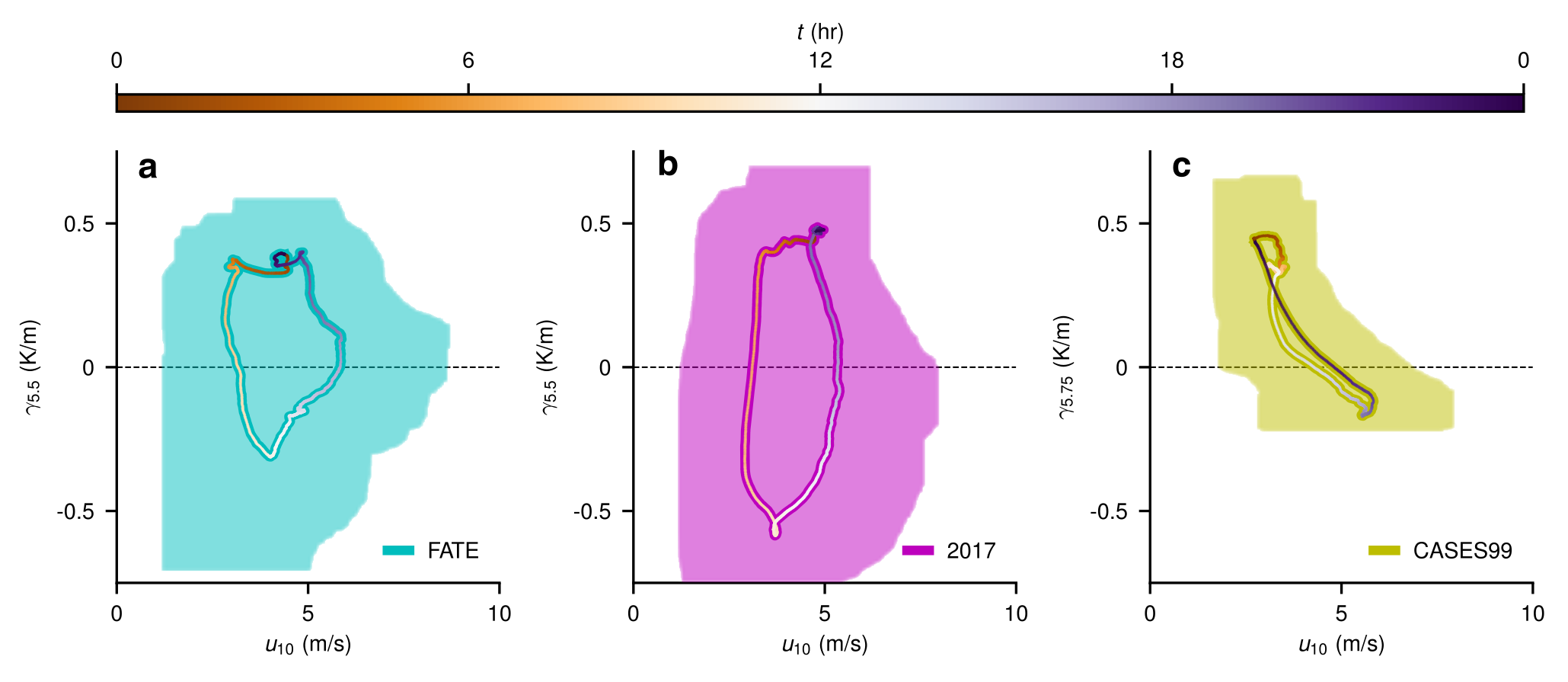}
\caption{Steady and unsteady daily ABL trajectories. Each plot shows average 1-hr smoothed daily trajectories colored by day-hour $t$ of 10-m wind speed $u_{10}$ horizontally and near-surface potential virtual lapse rate $\gamma$ (taken at 5.5 m at White Sands, New Mexico and 5.75 m near Leon, Kansas). Shaded regions surrounding the trajectories are the union of interquartile range rectangles for all times. (\textbf{a}) FATE measurements are taken in the 2018 windy season at the location in Fig. S1. (\textbf{b}) Measurements taken on the upwind Alkali Flat in the 2017 windy season (see Fig. S4). (\textbf{c}) CASES99 measurements are taken from the 55-m Main Tower in October 1999 near Leon, Kansas \cite{poulos2002cases}. We posit that (a-b) are unsteady ABLs and (c) is steady, by the presence and (near-)absence of an area inscribed by their trajectories, respectively. The weakly-forced ABL in CASES99 results in negligible hysteresis.}
\end{figure}

\begin{figure}
\centering
\includegraphics[width=0.64\textwidth]{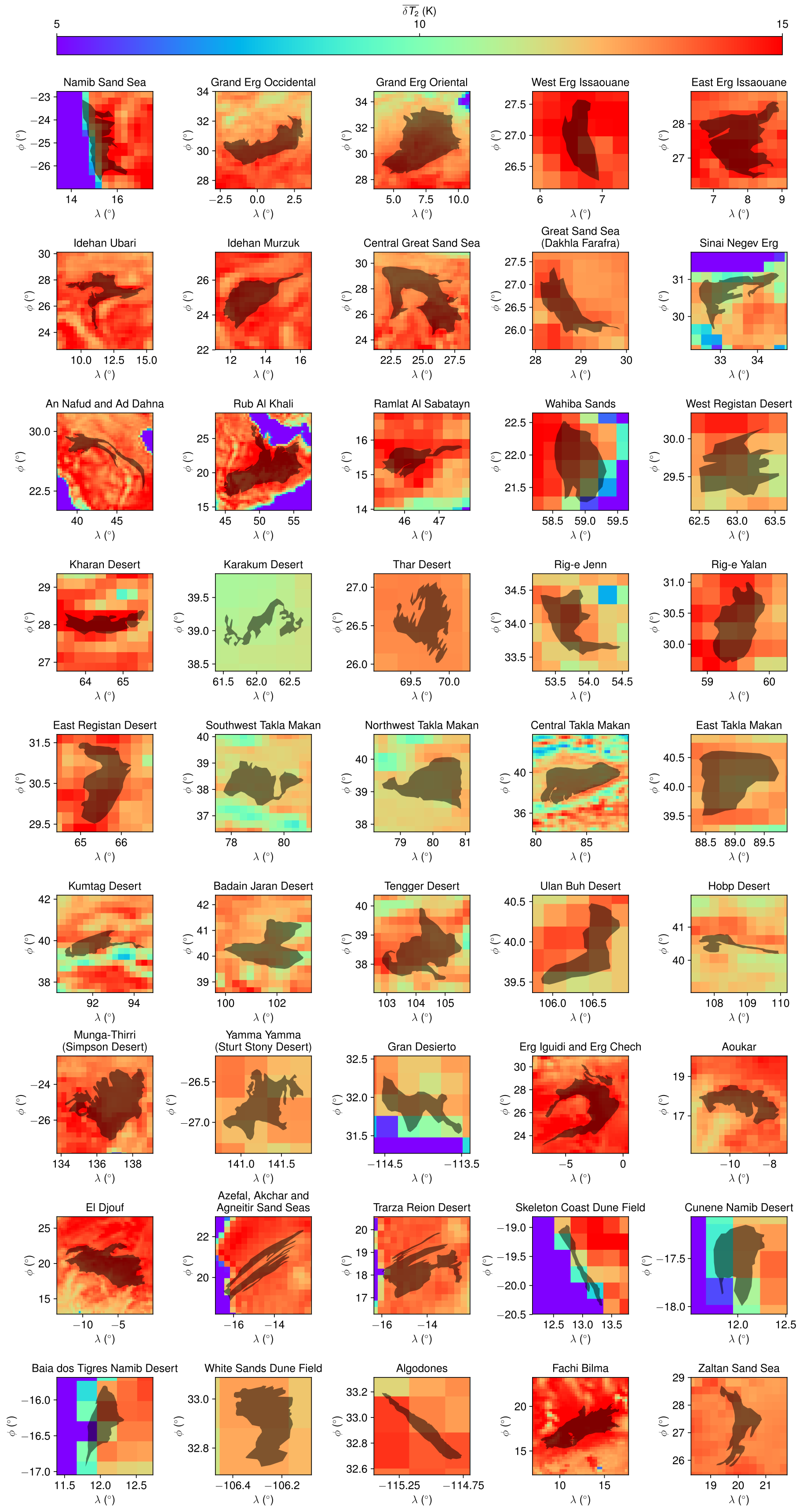}
\caption{Active sand seas on Earth. Shaded outlines of the 45 dune fields used in the global study overlaid on a map of 2008-2017 average of diurnal temperature range from ERA5 \cite{copernicus2017era5}. Names of each dune field are above each subplot, integer latitude $\phi$ and longitude $\lambda$ are marked on the axes, a common colorbar for each panel is presented at the top of the figure. Any ERA5 land grid-cells overlapping with the dune field outlines are included in their analysis (see \textit{Materials and Methods}).}
\end{figure}

\begin{figure}
\centering
\includegraphics[width=0.475\textwidth]{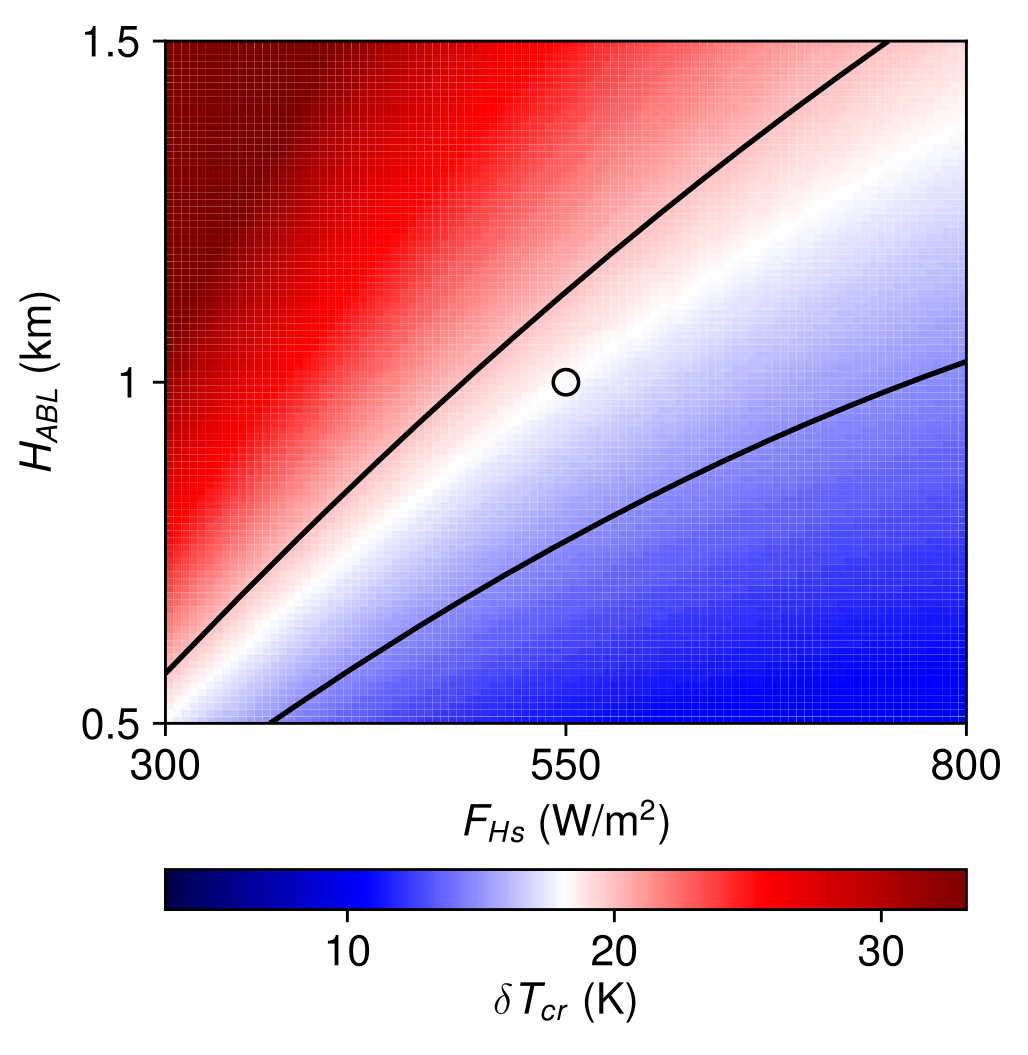}
\caption{Sensitivity of critical diurnal temperature range to ABL properties. A contour plot of the equilibrium amplitude of daily sinusoidal near-surface temperature change ($\delta T_{cr}$) from an ABL `slab model' \cite{garratt1994atmospheric} (see \textit{Materials and Methods}). Two free parameters, the surface sensible heat flux ($F_{Hs}$) and the ABL height ($H_{ABL}$) span the axes. The open circle in the center of the plot denotes the value of  $\delta T_{2,cr} = 18.1$ K used in Fig. 5a. Contours of $\delta T_{cr}$ as 15 K and 20 K are marked. The colorbar is centered on the value in the plot center.}
\end{figure}

\FloatBarrier

\movie{GOES-16 satellite training set for dust detection. Six segments of observations from the GOES-16 satellite corresponding to the training set for the machine learning dust detection algorithm (see \textit{Materials and Methods}). Each frame in this video (20 frames per second) is a single capture from the satellite ABI (local time in top left) which has a temporal resolution of 5 mins. On the left, a `True Color' image constructed from the 0.47 $\mu$m (blue), 0.64 $\mu$m (red) and 0.86 $\mu$m (green) bands is presented. On the right, a `split window' image is constructed from the residual of the 11.2 $\mu$m and 8.5 $\mu$m bands, which is commonly used to find clouds \cite{schmit2018applications}. Each video segment is titled on the top right, where each class in the machine (`Clear Sky', `Cloud Cover', and `Dust Emission') has 2 segments each that are numbered and continuous in time. Overlaid on both image constructions are semi-transparent RGB `False Color' masks output from the machine learning algorithm. This mask shows the probability prediction of machine learning algorithm for the current frame, where dust emission (red), clear sky (blue) and cloud cover (green) are the pure colors (Fig. S2).}

\bibliography{pnas-sample}